\documentclass[11pt]{article}
\usepackage{graphicx}
\begin{document}
\begin{center}
{\large {\bf Performance of An Axial Gas Ionization Detector}}\\
\vskip 1cm
S. Adhikari$^1$, C. Basu$^1$, C. Samanta$^{1,2}$, S. S. Brahmachari$^1$, 
B.P. Das$^1$ and P. Basu$^1$\\
\vskip 0.7cm
$^1$ Saha Institute of Nuclear Physics, 1/AF Bidhan 
nagar, Kolkata-700064, India.\\
$^2$ Physics Department, Virginia Commonwealth University, Richmond, 
VA 23284-2000, USA.\\
\vskip 1cm
\begin{abstract}
An axial gas ionization chamber has been fabricated for use as a $\Delta E$ 
detector in heavy ion induced nuclear reactions. Different operating parameters
 such as gas type, pressure, anode voltage and anode structures have been 
optimized. The transparency of the anode structure is observed to play an 
important role in improving the energy resolution of the detector.
\end{abstract}

\end{center}
\vskip 1cm

\section{Introduction}
Detection of charged particles provides important information 
about the 
reaction mechanisms induced by accelerated heavy ions. For light charged 
particles (LCPs) such as protons, deuterons, alphas solid-state silicon 
detectors provide excellent energy, timing and position resolution. 
Moreover, due to higher density of the detecting medium solid-state 
silicon detectors offer better opportunity to stop high energy charged 
particles. The solid-state detectors are also prone to permanent radiation 
damage and therefore inadequate for heavy ion detection. Common 
problems with highly ionizing particles for these detectors are pulse height 
defect and plasma delay. Corrections in the measurement are then required to 
get correct information about the energy and rise time of the interacting 
radiations.\\

In highly ionizing environment, on the other hand, gas detectors offer better 
options \cite{as92}. These detectors permit operation under high current rates 
and high dose environment and the gas can be recycled 
to maintain the purity of the detecting medium. Gas detectors for charged 
particles are primarily of two types: 1) the low pressure multiwire counters 
which show excellent position and timing resolutions \cite{ba02} and 
2) relatively high pressure 
ionization or proportional chambers for identification and energy measurement 
of various heavy ions emitted from a nuclear reaction.\\

Ionization chambers generally operate at the reduced electric field values of
$E/P$ $\sim$ 1 to 2 Volts/(cm$\times$Torr) where $E$ is the applied electric 
field and $P$ is the gas pressure. In a typical design of an ionization chamber
\cite{kn00}, a transverse field is applied. This field arrangement proves to be
disadvantageous as the associated pulse height becomes a function of position
where the incident particle impinges \cite{kn00}. Although, this problem is 
solved by the introduction of a Frisch grid to the ionization chamber, in an 
axial field configuration a Frisch grid is not always useful.\\ 

The present work describes the performance of a $\Delta E$ ionization chamber 
working in the axial field configuration.  
The advantage of using an axial field in a $\Delta E$ detector over the more usual 
transverse field has 
been discussed in ref [4,5]. Since then there has been no studies 
or development on this form of gas detector. Extensive studies on the different
parameters are sparse in the literature. In this paper we have made a detailed 
study on the different operating parameters of our detector. The gas 
pressure, bias voltage and window foil thickness of the chamber have been optimized.
Earlier studies [4,5] have given little or no importance to the anode 
structure on the detector performance. A parallel wire anode structure has been used
for the first time to achieve very high transparency and its possible effects on the
energy resolution have been addressed.\\

\section{Construction of the detector}

\begin{figure}
\centering
\includegraphics[width=4.5in]{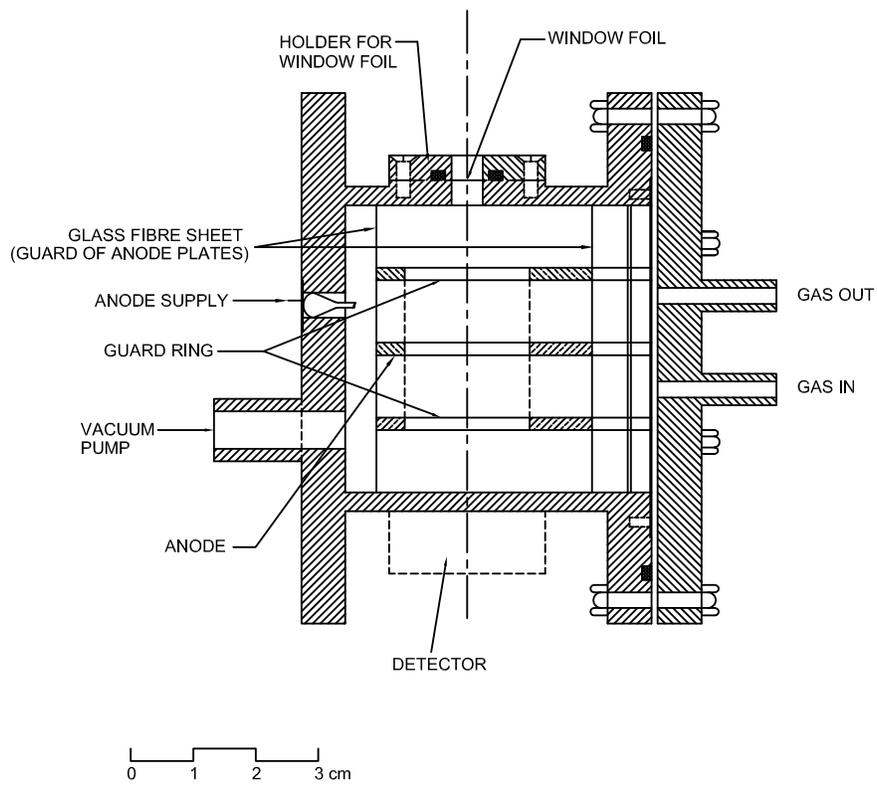}
\caption{Cross-sectional view of the gas ionization chamber}
\label{fig_cons}
\end{figure}
Fig.1 shows the construction of the detector used in the present studies. 
The body of the detector is 
made of aluminium with an active area of 45$\times$45 mm$^2$. The window 
is an aluminized mylar foil mounted by grease (non outgassing at pressures of
at least 10$^{-6}$ Torr) and held by an O-ring in a brass flange. The 
window foil is maintained at ground potential by contact with the screws of the
brass flange. 
The anode structure is located in the middle of the chamber and is mounted on a 
brass plate of active area 34$\times$34 mm$^2$ and thickness 2 mm. The plate is 
provided with a hole of diameter 20 mm. The anode structure is
in the form of conducting parallel thin wires or mesh and is mounted on a 
1.5 mm thin copper clad G10 board (PCB) of area 30$\times$30 mm$^2$ with a 
square 
hole 20$\times$20 mm$^2$. The anode assembly is fixed by teflon screws with 
the central brass plate which is stepped down by the PCB thickness.
Different thin wire structures (parallel or crossed) in the anode are used
to study its effect on the energy resolution of the detector. In order to maintain 
the uniformity of the electric field gradient along the incident particle path 
two additional brass plates are kept at half the anode voltage at 10 mm on 
either side of the anode. In this way the separation between the plates is made 
smaller compared to their lengths so that the effect of non-uniform electric 
field can be ignored. The brass plates are of the same thickness and 
area as that of the central plate except the hole is kept empty. The 20 mm 
hole of the guard plates and that of the anode are coaxial with the 5 mm hole 
of the window. The separation of the first guard plate from the
window and the last guard plate to the exit is 9.5 mm. Since the detector will 
be ultimately used as a $\Delta$$E$ detector the anode is centrally located so 
that the active volume is extended on both sides of the anode plane.\\ 

The voltage to the anode is provided by a kovar seal and the divided 
voltage to the plates is provided by a breeder 
(resistive) circuit placed inside the chamber. The plates with applied voltage 
are isolated from the ground by G10 spacers which cover the inner walls 
perpendicular to the window and exit side. The body of the detector at the 
window (including the window foil) and the exit side is at ground.  
This ensures the field lines along
the axis of the chamber. The DC current path and other details of the 
equivalent circuit describing the resistive gradient is shown in fig.2. 
Calibrated resistances of value 50 M$\Omega$ are used to make the resistive 
gradient. 
\begin{figure}
\centering
\includegraphics[width=4.5in]{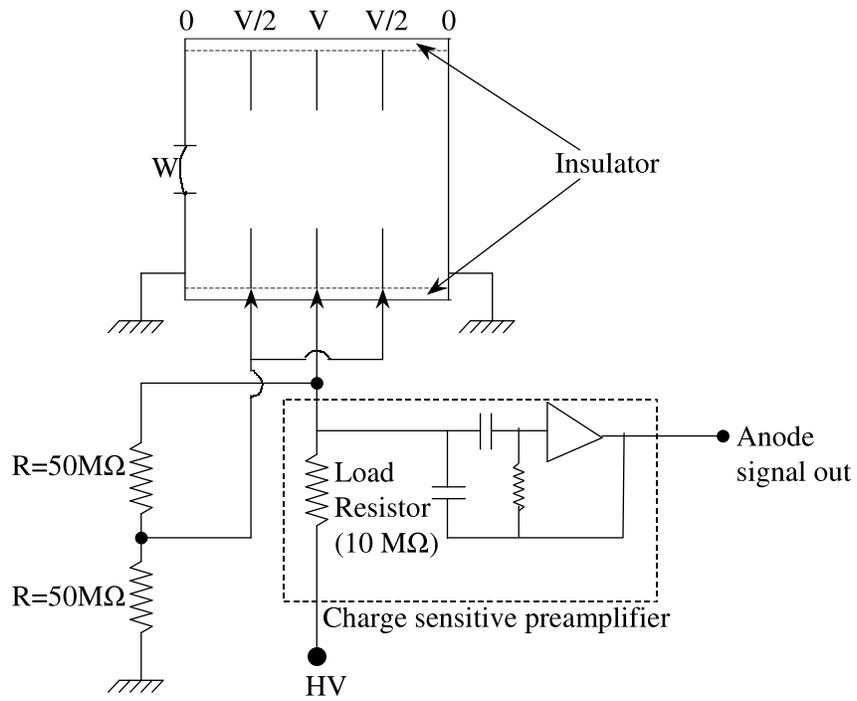}
\caption{Equivalent circuit describing the resistive gradient and 
all the ground paths. The details of the preamplifier (Ortec 142IH) circuit 
are available in the manual.}
\label{fig_cir}
\end{figure}
Resistive noise in our case was negligible. The noise level at the 
preamplifier stage was within 15 mV. One side of the detector is flanged 
with a viton 
O-ring for dismantling and testing. The gas inlet-outlet and provision for 
evacuation of the detector are shown in the figure 1.
The system is pumped down to 10$^{-3}$ Torr vacuum 
and charged with gas at the desired pressure.
The detector is now ready to be tested.\\  

\section{Experiment and discussion of results}
 
The detector performance was tested with a $^{252}$Cf 
$\alpha$-source. 
The $\alpha$-source was collimated to fall on the entrance window which is 5 mm 
in diameter. The collimation and thickness of the window foil plays an 
important role on the resolution of the detector. We used a collimation 
of 1 mm and mylar foils for the window of two different thickness. The 
detector was filled with gas between 60-600 Torr pressure. Two different
gases Ar(90$\%$)-CH$_4$(10$\%$) and isobutane were used to compare the performance 
of the chamber in each case. The output of the 
detector was fed to a charge sensitive preamplifier (ORTEC 142IH) and to a 
spectroscopy amplifier (ORTEC 672) and finally to a 2K ADC MCA (ORTEC 
maestro-32).  
Fig.3 shows the acquired energy-loss spectrum of $\alpha$ particle
(84$\%$ 6.12 MeV and 15.7$\%$ 6.08 MeV) 
from a collimated $^{252}$Cf source of strength 10 $\mu$Ci.\\
\begin{figure}
\centering
\includegraphics[width=4.5in]{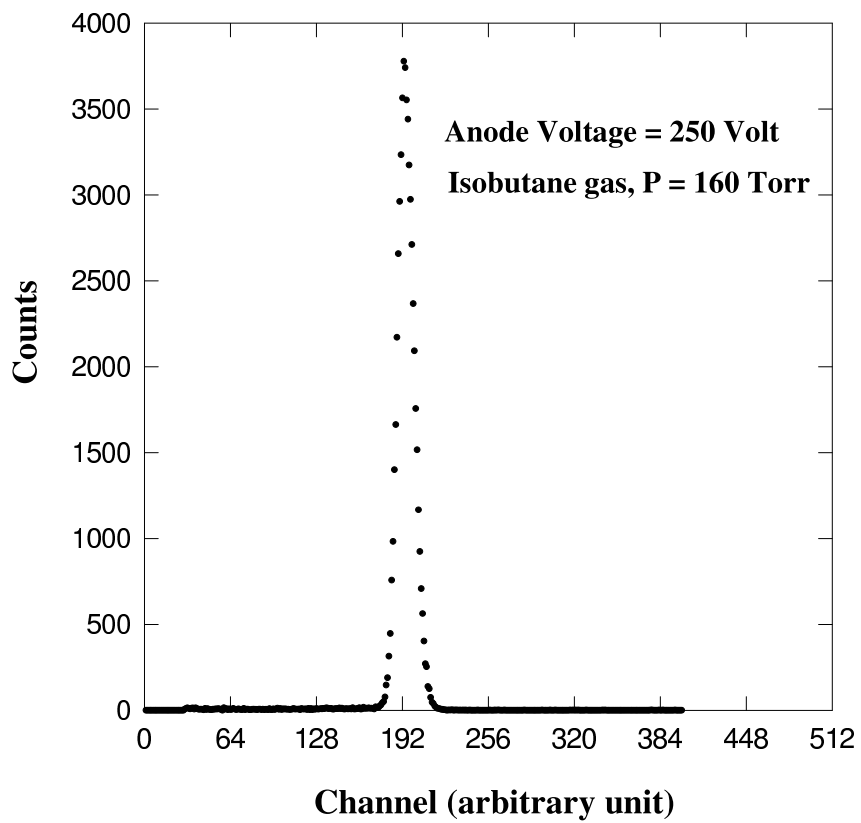}
\caption{The energy-loss spectrum of $\alpha$-particle (84$\%$ 
6.12 MeV and 15.7$\%$ 6.08 MeV) from a collimated $^{252}$Cf source.
The energy resolution is 6.8$\%$ FWHM.}
\label{fig_alpha}
\end{figure}
\\
\begin{figure}
\centering
\includegraphics[width=3.5in]{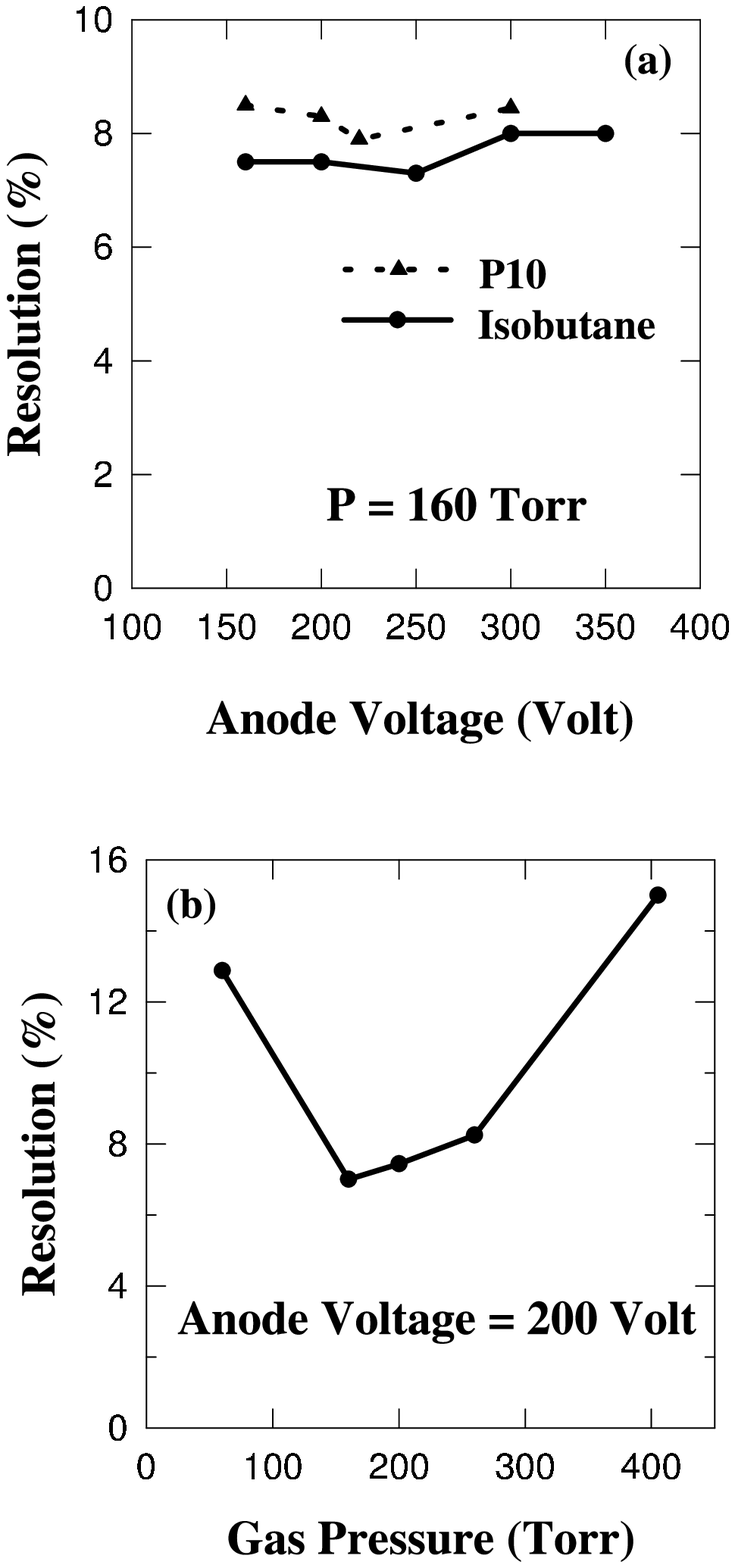}
\caption{Variation of energy resolution with {\bf (a)} anode voltage at a 
fixed gas pressure (160 Torr) for two different types of gas Isobutane and 
Ar(90$\%$)-CH$_4$(10$\%$) (P10) and {\bf (b)} with gas pressure at a fixed anode 
voltage (200 V).}
\label{fig_para}
\end{figure}
 
The different parameters of the detector were optimized to obtain the best
possible energy resolution with the collimated alpha spectrum. 
The energy calibration for the $\Delta E$ detector was performed with
a $^{241}$Am and $^{252}$Cf 
$\alpha$-particle (5.485 MeV and 6.12 MeV respectively) energy loss  
and its correlation with the pulse heights. 
The stopping power of $\alpha$ particles in gases are well accounted by the 
Bethe-Bloch formula \cite{pa66}. The mean energy loss (centroids) in the gas and
mylar window were thus determined from Bethe-Bloch formula (code SRIM 
\cite{zi66}) with the density of gas scaled by 
the ratio of desired to the normal gas pressure (760 torr). 
The centroid of the data was obtained by an in-built fitting software 
of the MCA.
This method of energy calibration for $\Delta E$ detectors using simulated 
energy loss have been adopted in studies with solid-state detectors \cite{al00}.
Two different window 
foils of thickness 1.5 and 8 $\mu$m were examined. For the thicker
foil the resolution varied between 9-13$\%$ FWHM whereas for the 1.5 $\mu$m
window the resolution was much better (6-8$\%$ FWHM). A possible reason for this
is that higher straggling effects in the thicker window foil degrades the 
resolution of the detector.
As the area of the window aperture is small 
noticeable effect of window deformation due to pressure differences on the 
either side of the window was not observed. 
As far as the working gas is concerned, better performance was obtained with 
isobutane in 
comparison to Ar(90$\%$)-CH$_4$(10$\%$). This is depicted in fig. 4(a). A reason
for this may be due to the higher energy loss per unit  pressure in isobutane 
\cite{ja83}.
The optimum gas pressure was 
found to be at 
160-200 Torr and bias voltage 200-250 Volts. We show in fig.4 (b) the region
of optimum gas pressure (160-250 torr) at 200 Volts bias. At lower gas 
pressures (60 - 100 Torr) 
the recombination effect is overcome earlier though the resolution is poorer 
than at higher pressures. The statistical fluctuation in the number of
electron ion pairs created is larger at lower pressure
due to lower energy loss (hence poorer resolution). At higher pressure the 
resolution is mainly limited due to increased recombination [4,10].
The energy loss of the alpha particle in the gas 
and FWHM of the observed peak in energy unit is given in table 1 at 
anode voltage, $V=200$Volt and shaping time$=3 \mu s$. 
The operating reduced electric field ($E/P$) was found to be between 
1-2 Volts/(cm$\times$Torr). 
A gaussian shaping for optimum signal-to-noise ratio was used and the optimum 
amplifier shaping time was found to be 
3$\mu$s (charge collection time for our detector is about 400 $ns$). 
The resolution degraded at higher and lower shaping time. The detector could 
handle about 500 counts per second with good resolution.\\ 
\begin{figure}
\centering
\includegraphics[width=4.5in]{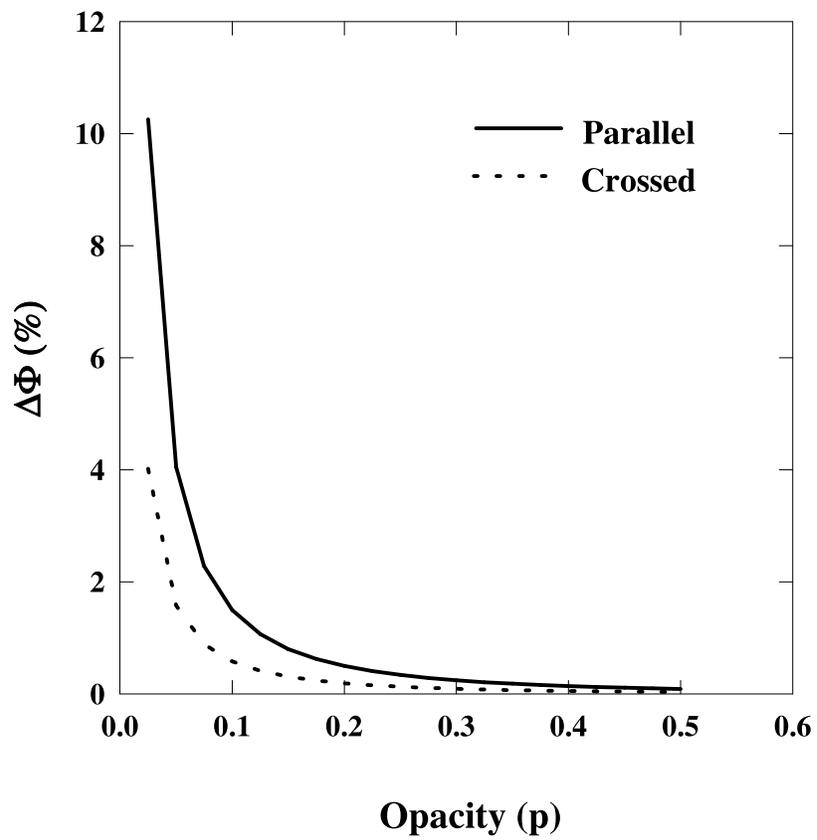}
\caption{Plot of variation of percentage potential shift per unit 
field difference ($\Delta$$\Phi$) with opacity($p$) for parallel and crossed 
wire mesh.}
\label{fig_ef}
\end{figure}

\begin{table}
\renewcommand{\arraystretch}{1.3}
\caption{Energy loss and resolution in keV are given at different gas 
pressures for Isobutane and P10 gases.}
\label{table_example}
\centering
\begin{tabular}{|c||c||c||c||c|}
\hline
 & \multicolumn{2}{c||}{Isobutane} & \multicolumn{2}{c|}{P10}\\
\cline{2-5}
Gas Pressure & Energy Loss & FWHM(expt) & Energy Loss & FWHM(expt)\\
(Torr) & (keV) & (keV) & (keV) & (keV)\\
\hline
60 & 365.3 & 47.03 & 261.3 & -\\
\hline
160 & 973.6 & 68.15 & 697.1 & 55.77\\
\hline
200 & 1217.6 & 90.56 & 871.3 & 74.06\\
\hline
260 & 1582.8 & 130.48 & 1132.8 & 107.61\\
\hline
405 & 2466.0 & 369.90 & 1764.4 & -\\
\hline
\end{tabular}
\end{table}

The effect of the anode structure 
on the energy resolution has not been addressed by earlier 
workers [4,5]. In ref \cite{ba89} an electro-formed nickel mesh 
was used
with high transparency (97$\%$). But no justification was given for choosing 
such an anode structure. F.H. Read et al \cite{re98}, have made an extensive 
computational study about the 
electrostatic problems involving a mesh sandwiched between plates on either 
side kept at different voltages. The effect of a mesh made of a) parallel ultra
thin round wires and b) crossed round wires have been studied. 
According to this work, the potential on the mesh is modified depending 
on the structure and transparency of the mesh.
The fluctuation of potential on the mesh ($\phi_m$) is defined as
\begin{eqnarray}
\phi_m&=&\Delta {\cal{E}} \Delta \phi\\
\Delta \phi &=&s\chi_m
\end{eqnarray}
where $\Delta \cal{E}$ is the difference in electric fields on either side of 
the mesh, $\Delta \phi$ denotes the potential shift per unit field difference, 
$\chi_m$ is  a dimensionless parameter 
that depends on the transparency (opacity) of the mesh and $s$ is the 
separation between any two adjacent wires in the mesh \cite{re98}. 
The percentage variation of potential
shift per unit field difference ($\Delta \phi$ $\%$) for crossed and parallel
wires with opacity ($p$) is displayed in figure 5.  
It is clearly seen that with very high transparency (small opacity) 
$\Delta \phi$ is very large and the mean voltage on the mesh will
shift considerably even if $\Delta \cal{E}$ is small.
(For the present electrode configuration the magnitude of the electric field 
is same but changes sign on either side of the anode.)
At the same transparency however this shift is reduced by a factor of 2 in 
comparison
to a mesh of parallel wires (figure 5). This is due to the reduction of the
parameter $\chi_m$ by this factor in the latter case \cite{re98}.
It is therefore interesting to study how the fluctuation of potential (if any) and
structure of the anode affect the resolution of the detector.\\

Anode structures made of parallel and crossed ultra thin wires 
were used in the present ionization detector. 
For the crossed anode structure an electro-formed 
Nickel mesh acquired from Precision e-forming, USA 
with transparency of 89$\%$ was used. For this mesh the expected  potential shift
is only 1$\%$ for our detector configuration. The best resolution that 
obtained with this anode structure was about 7.5$\%$ FWHM. 
The anode structure made of parallel wires was self-fabricated with gold 
plated platinum wires. The wire diameter was kept 50$\mu$m and wire spacing 
2 mm. The transparency in this case was 97.5$\%$.  A resolution of 7.3$\%$ 
FWHM was obtained with this anode structure. Though for the latter
anode structure $\Delta \phi$ could be as high as 10$\%$  we 
did not observe any significant effect of this increase on the resolution. 
Instead the effect of transparency could be seen through the improvement 
in resolution with
increase in transparency. The reason for better resolution for more
transparent electrodes is due to the reduced input capacitance of the
preamplifier. 
 The opacity of a crossed mesh is roughly twice 
than that with parallel wires for the same ratio of wire-diameter to 
wire-spacing. Therefore it is more convenient to work with parallel wires 
that can be fabricated to very high transparency. In order to test the effect 
of transparency on resolution we used another anode structure where the wire 
diameter was taken to be 12.5 $\mu$m, keeping the wire spacing to be 2mm as 
before. The transparency for this case was even higher (99.4$\%$). Though the 
expected potential fluctuation on the wires is very high in this case we found 
an improvement in resolution, which is found to be within 7.0$\%$ FWHM.\\ 

In the extreme case of a blank anode i.e. the anode frame made 
of brass plate with
 the 20 mm hole at the center was used without any mesh or wires. Here in this 
case however no improvement of resolution was observed. 
This structure also 
required a higher amplifier shaping time (6 $\mu$s) for the best possible 
resolution (8$\%$).  This is because with the blank hole anode the charge 
collection is not as efficient as with an anode with conductors in the hole.
The use of a highly transparent mesh as anode is therefore
suitable for improving the resolution. This can be achieved more easily
 with parallel
arrangement of anode wires rather than with a crossed structure. In all the
measurements, fresh gas was used
for each different anode structure so that the relative error between any
 two measurement (due to gas impurity) is negligible.\\

\begin{figure}
\centering
\includegraphics[width=4.5in]{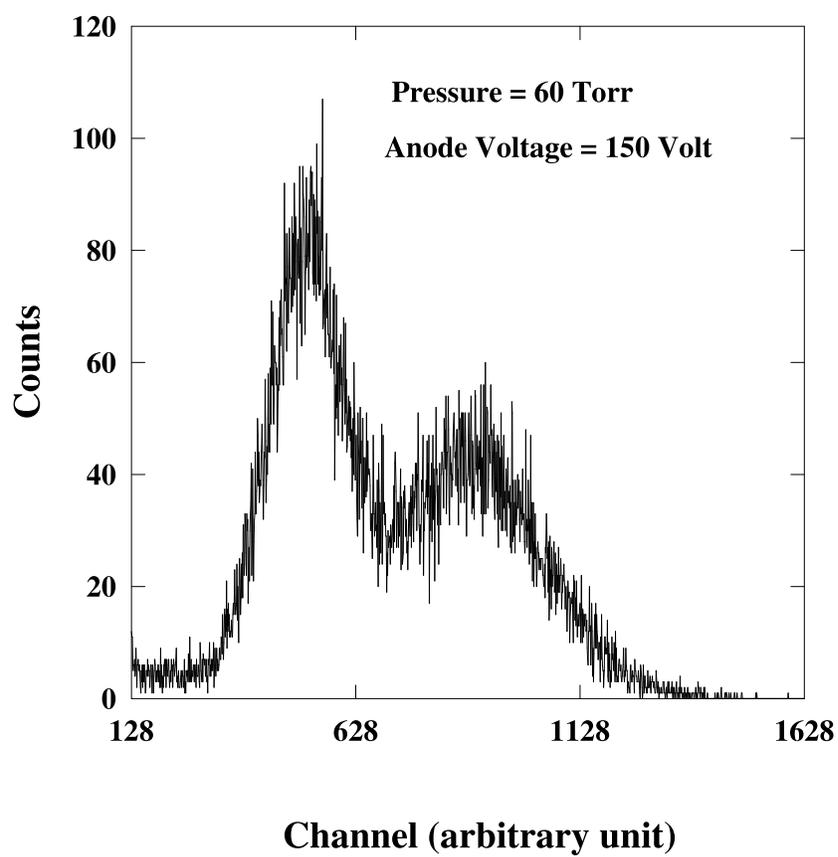}
\caption{The energy-loss spectrum of spontaneous fission fragments 
from $^{252}$Cf source. The heavy fragments on the left have higher 
intensities than that of lighter fragments.}
\label{fig_fission}
\end{figure}

Performance of the detector for heavy ions were studied 
by recording the $\Delta$$E$ 
spectrum of the spontaneous fission fragments from $^{252}$Cf. The acquired
spectrum is displayed in fig.6. The light (on the right) and heavy fragments 
(on the left) are observed 
to be reasonably separated. The optimum anode structure was the one with
12.5$\mu$m parallel thin wires and gas pressure at  60 Torr.
At higher pressures the separation degraded possibly due to increased 
recombination. The peak to valley ratio was seen to improve with collimation
of the source. The fission $E$ spectrum parameters for a solid-state 
detector described in ref \cite{kn00} are evaluated
for the present detector and depicted in table 2. The  $^{252}$Cf fission
spectrum has been studied in \cite{aj72} by a gas $\Delta E$ chamber where
the best peak to valley ratio is 2.8. However, in \cite{aj72} the source is placed
inside the detector volume and the window effect is avoided. In \cite{sa75} 
$^{252}$Cf spectrum has been measured by a gas $E$ detector and the peak to valley
ratio for light and heavy fragments are quoted as 2.25 and 2.07 respectively. 
These parameters in the present case are thus within reasonable 
limit.
It should be however noted that the intensities of the light and 
heavy fragments interchange in
the recorded $\Delta$$E$ spectrum in comparison to a typical $E$ spectrum of 
the fission fragments. The heavy fragments (left) have higher intensity 
than that of the lighter fragments. This is owing to increased 
energy straggling of the lighter
fragment than the heavier one as studied in ref \cite{sy71}. Similar shift
in the heavy and light group has also been observed by \cite{aj72} but at much higher
pressures. In-beam studies 
with a gas $\Delta E$ detector in conjunction with a stopping solid-state
detector were not done as they have been pursued in detail by several workers
[15-19].\\

\begin{table}
\renewcommand{\arraystretch}{1.3}
\caption{Parameters of the $^{252}$Cf $\Delta E$ fission fragment spectrum. The
definition of the different parameters are same as in ref \cite{kn00}.}
\label{table_example}
\centering
\begin{tabular}{|c||c|}
\hline
Spectrum Parameter & Values\\
\hline
$N_H/N_V$ & 3.8\\
\hline
$N_L/N_V$ & 2.2\\
\hline
$N_H/N_L$ & 1.7\\
\hline
$\Delta$$L/(L-H)$ & 0.3\\
\hline
$\Delta$$H/(L-H)$ & 0.27\\
\hline
$(H-HS)/(L-H)$ & 0.53\\
\hline
$(L-LS)/(L-H)$ & 0.69\\
\hline
$(LS-HS)/(L-H)$ & 2.23\\
\hline
\end{tabular}
\end{table}

There are however certain limitations on the detector performance.
For example, the charge induced due to moving electrons ($q_{ind}^{(-)}$)
and positive ions ($q_{ind}^{(+)}$) is given by the Shockley-Ramo Theorem
[3,20,21] as
\begin{eqnarray}
q_{ind}^{(-)}&=&-\frac{e}{W}\int_{0}^{d}\frac{dE(z)}{dz} [\phi_{w}(d)-
\phi_{w}(z)]dz \\
q_{ind}^{(+)}&=&-\frac{e}{W}\int_{0}^{d}\frac{dE(z)}{dz} [\phi_{w}(z)-
\phi_{w}(0)]dz 
\end{eqnarray}
where $e$ is the electronic charge, $W$ is the mean energy for ionization,
$\frac{dE(z)}{dz}$ the stopping power of the incident particle in the gas
and $\phi_{w}(z)$ is the weighting potential in the present case
at any point $z$ along the 
particle track, $d$ being the anode-cathode separation. If the particle
stops inside the detector active length the upper limit in the integrals
should be replaced by the range of the particle.\\ 

The total induced charge is a result of the motion of all the ionizations
that occur along the track between the cathode and anode. 
As can be seen from equations (3) and (4),
in the electron sensitive operation 
\cite{kn00} of the detector, the anode signal is not proportional to the 
energy deposited by the particle (due to the $z$ dependence of $\phi_{w}$).
This problem can be reduced if a Frisch Grid 
is placed between the cathode and anode. 
The weighting potential is now suppressed between the cathode and grid 
($\phi_w=0$ for $0<z<b$) so that equation (3) reduces to
\begin{eqnarray}
\nonumber q_{ind}^{(-)}&=&-\frac{e}{W}\int_{0}^{b}\frac{dE(z)}{dz} [\phi_{w}(d)-
\phi_{w}(z)]dz \\
&-&\frac{e}{W}\int_{b}^{d}\frac{dE(z)}{dz} [\phi_{w}(d)-\phi_{w}(z)]dz
\end{eqnarray}
The proportionality to energy loss is thus ensured in the region between cathode
and grid. However, the source of non-uniformity (second term of (5)) cannot 
be completely eliminated in an axial mode but can be reduced
by keeping the separation between grid and anode small. In this work
we have used a very simple electrode structure keeping the holes in
 the guard plates empty. The weighting potential
for the present electrode configuration ($b/d=0.5$) is calculated by using 
the formalism of ref \cite{ja89} (subject to the boundary conditions in the 
present case) and depicted in fig.7. The weighting potential along the axis 
of the detector is plotted where the distortion in the region between the
cathode and the guard plate 
due to the hole in the plate is maximum. This effect reduces as one 
moves away from the axis. The effect of a guard plate without a 
hole which entirely suppress $\phi_w$ between the cathode and grid is shown 
for comparison. In the practical case a transparent mesh has to be used. 
However use of additional mesh besides the
anode may add to a background in the detector spectrum due to unwanted 
scattering
of the incident particles \cite{zu82}. Thus in an axial chamber the
Frisch grid is not completely effective as in a transverse field chamber.\\

\begin{figure}
\centering
\includegraphics[width=4.5in]{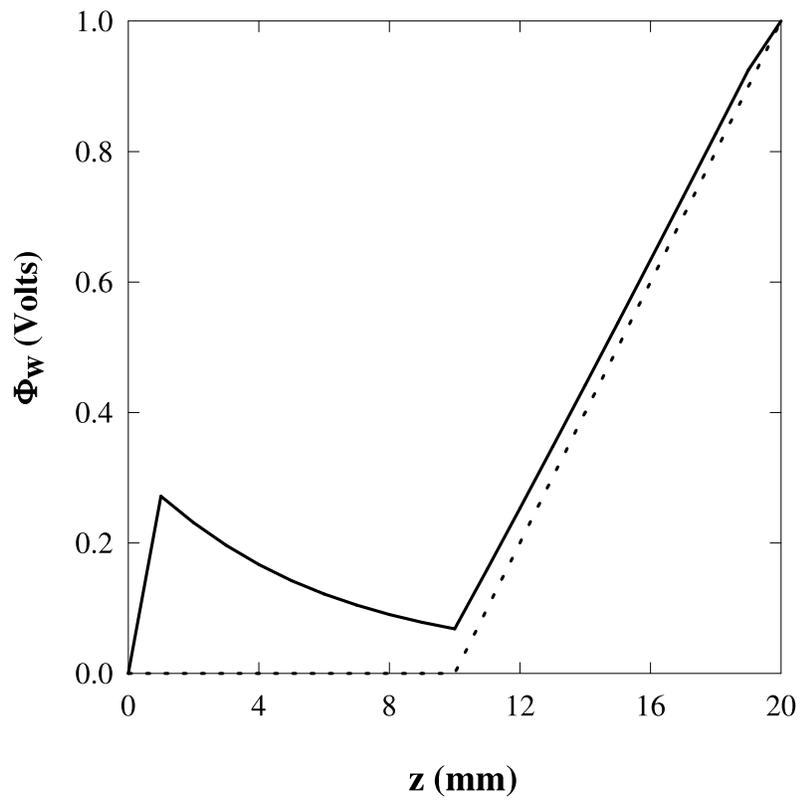}
\caption{Plot of weighting potential $\phi_w$ in the present electrode 
geometry (solid line) along the axis of the detector. The dotted line 
repesents the weighting potential for the case of guard plate with no hole.} 
\label{fig_fission}
\end{figure}

If however the Bragg-curve of the particle is approximately constant over the 
detector length, equations (3) reduces to
\begin{equation}
q_{ind}^{(-)} \approx - \frac{e}{W}S\int_{0}^{d} [1-
\phi_{w}(z)]dz 
\end{equation}
where $S$ is the constant stopping power.
The result of the integration will always be a function of $d$ (i.e, 
independent of $z$) irrespective of the form of $\phi_w(z)$.  
Thus the measured signal will be proportional to the energy lost ($Sd$) 
by the particle in the detector active length.
In the present work the Bragg-curve for a 6.12 MeV $\alpha$
particle and the $^{252}$Cf fission fragments are almost constant over the 
cathode-anode gap (20 mm) and so the detector functions satisfactorily as
a $\Delta E$ device.

\section{Summary and conclusion}
We have fabricated a gas ionization chamber working in the axial mode charge 
collection configuration using parallel plate geometry. The different 
parameters like the gas type and pressure, anode voltage and anode structure 
have been optimized.
In particular the effect of anode structure has been studied. A mesh with
parallel array of round wires was found to give better energy resolution. A higher
transparency is found to give better resolution, although a larger fluctuation
of the mesh voltage was associated with it. Therefore, a parallel array of 12.5 
$\mu$m round gold plated wires will be used in future to get better resolution.

\section{Acknowledgement}

The author(S.A.) would like to thank Ms. S. Bhattacharyya, Prof. S. Saha, Mr.
Dulal Ghosal, Mr. S. Chakraborty and SINP workshop for their help at different 
stages of this work.\\

\end{document}